\documentclass[11pt]{article}
\usepackage{epsf}
\newlength{\normalarraycolsep}
\newlength{\normaltabcolsep}
\newcommand{\be}{\begin{equation}}
\newcommand{\ee}{\ \ \ \end{equation}}
\newcommand{\bea}{\setlength{\arraycolsep}{0.4\normalarraycolsep} 
                  \begin{eqnarray}}
\newcommand{\eea}{\ \ \ \end{eqnarray}\setlength{\arraycolsep}
                  {\normalarraycolsep}}

\newcommand{\arcsinh}{\mathop\mathrm{arcsinh}\nolimits}

\newcommand{\arctanh}{\mathop\mathrm{arctanh}\nolimits}
\newcommand{\sign}{\mathop\mathrm{sign}\nolimits}

\newcommand{\cn}{\mathop\mathrm{cn}\nolimits}

\newcommand{\fpb}{\bar{f}_\mathrm{PB}}

\newcommand{\fm}{\phi_\mathrm{m}}
\newcommand{\dd}{\mathrm{d}}
\newcommand{\e}{\mathrm{e}}

\newcommand{\fl}{\phi_<}
\newcommand{\fg}{\phi_>}
\newcommand{\fii}{\phi_\mathrm{i}}
\newcommand{\xii}{x_\mathrm{i}}

\begin{document}
\title{\bf Electrolytic depletion interactions}
\author{M. N. Tamashiro and P. Pincus\\
{\it Materials Research Laboratory, University of California
at Santa Barbara}\\
{\it Santa Barbara, CA 93106-5130, USA}\\
e-mail: mtamash@mrl.ucsb.edu, fyl@physics.ucsb.edu}
\maketitle
\begin{abstract}
We consider the interactions between two uncharged planar macroscopic 
surfaces immersed in an electrolyte solution which are induced by interfacial
selectivity. These forces are taken into account by introducing a depletion 
free-energy density functional, in addition to the usual mean-field 
Poisson-Boltzmann functional. The minimization 
of the total free-energy functional yields the density profiles of the microions 
and the electrostatic potential. The disjoining pressure is obtained by differentiation
of the total free energy with respect to the separation of the surfaces, holding the 
range and strength of the depletion forces constant. 
We find that the induced interaction between the two surfaces is always 
repulsive for sufficiently large separations, and becomes attractive 
at shorter separations. The nature of the induced interactions changes from
attractive to repulsive at a distance corresponding to the range of the 
depletion forces.
\end{abstract}

\section{Introduction}

Electrostatic interactions often play an important role 
in a variety of different systems, ranging from biological 
membranes to chemical industrial paint ingredients. In some cases, 
it provides the underlying mechanism for the stabilization of 
mesoscopic systems against flocculation and precipitation.
When two macroscopic charged surfaces approach one another, 
the result is \textit{usually} a repulsive force, which inhibits 
a further approach. For two flat charged plates, this effect 
can be understood in a physical picture in terms of the osmotic pressure 
generated by the difference of the ion concentration in the region between the 
two approaching surfaces and the electrolyte-reservoir concentration.
On the other hand, attractive interactions, which lead to 
aggregation or fusion, are sometimes a desirable feature. This is the case, 
for example, in the adhesion and fusion of vesicles and membranes 
or in environmental sewage treatment. Furthermore, some 
experiments\cite{ise,bloomfield,podgornik,crocker,larsen} and 
simulations\cite{guldbrand,gronbech1,gronbech2,allahyarov} indicate that, 
for small separations and high surface-charge densities, two like-charged 
polyions can indeed attract. 

From the theoretical point of view,  
several distinct mechanisms leading to attractive 
interactions have been proposed, which are based on 
charge fluctuations\cite{spalla,pincus,levin}, 
strong positional charge correlations\cite{rouzina,shklovskii,arenzon},
anisotropic hypernetted chain calculations\cite{kjellander} or 
strong bulk-counterions correlations\cite{stevens,diehl}. 
Very recently a unified treatment taking into account 
quantum fluctuations and structural correlations of the Wigner crystals 
formed by the condensed counterions onto the charged surfaces 
 has been proposed\cite{lau}.
Although the bare Coulomb force between two macroscopic surfaces 
is always repulsive, correlations and/or fluctuations can 
induce attractive interactions, which occasionally may overcome the electrostatic
repulsion between the two equally charged surfaces. 
Correlations, which are entirely neglected within the 
mean-field Poisson-Boltzmann (PB) approximation 
(Gouy-Chapman theory\cite{gouy,chapman}), are believed to be 
\textit{essential ingredients} for the appearance of attractive interactions.
Thus, most proposed mechanisms which lead to attraction always include a 
non-mean-field effect.
 
In this work we propose a new mechanism for attraction between 
two identical plates. In contrast to the previous theoretical 
pictures, this mechanism is entirely at the mean-field level.
However, non-pure electrostatic forces are taken into account
by including depletion forces --- for example, those associated
with finite ionic radii --- acting on
one of the ion species surrounding the plates. 
For simplicity, we consider the case in which the identical 
plates are uncharged and infinitely large. By considering \textit{uncharged}
plates we can discern the effect of the depletion forces separately from 
the usual electrostatic mean-field repulsion, which indeed turns out to be 
entropic and not strictly electrostatic. 
If we treat the \textit{surface-charged} case, we are not able to
separate the two contributions. Due to the simplicity of the 
model, it is possible to derive explicit, analytical expressions for 
all thermodynamical properties, including the disjoining pressure.

The remainder of this paper is organized as follows. In Section 2
the model is introduced and the general equations are obtained.
Section 3 is devoted to solving the generalized PB equations 
for the non-overlapping regime, when the depletion zones associated 
with the two plates do not overlap. The solution to the generalized 
PB equations for the overlapping regime, when the depletion zones associated 
with the two plates do overlap, is obtained in Section 4. Some 
concluding remarks are presented in Section 5. The closed analytical 
expression for the disjoining pressure is obtained in the Appendix.   

\section{Definition of the model}

We shall consider two uncharged macroscopic surfaces immersed in a symmetric 
1:1 electrolyte within mean-field theory. The system is modeled by two planar,
infinitely thin rigid and uncharged surfaces, separated by a distance $h$,
in contact with a monovalent salt reservoir of bulk concentration $n_0$. 
A Cartesian coordinate system is chosen so that the surfaces are located 
at the $x=\pm h/2$ planes, in such a way that the $x$ axis is perpendicular
to the surfaces. At the mean-field level the microions 
are treated as an inhomogeneous ideal gas, with local number densities 
$n_+(x)$ and $n_-(x)$ for the positive and negative ions, respectively. 
We assume that, due to some \textit{non-electrostatic depletion} mechanism,
these local densities become inhomogeneous in the region close to the 
infinite plates. These inhomogeneities are governed by the reduced 
(total) free-energy functional (per unit area), $\bar f=\beta f$, where 
$\beta=1/k_\mathrm{B}T$,  
\be
\bar{f}= \bar{f}_\mathrm{depletion}+\fpb ,\label{eqn:totalfreeenergy}
\ee
which we split into two terms. 
The first term of (\ref{eqn:totalfreeenergy}) corresponds to a 
\textit{non-electrostatic depletion} free energy (per unit area),
\be
\bar{f}_\mathrm{depletion}=\epsilon 
\int\limits_{-\infty}^{\infty} \dd x\, n_+(x) \left[
 w_s\left(x+\frac{h}2\right)
+ w_s\left(x-\frac{h}2\right)\right],
\ee
where $\epsilon$ is a depletion-strength parameter and 
$w_s(\xi)$ can be considered a normalized external (non-electrostatic) 
potential with a finite short range, $s$. This term breaks the 
original degeneracy between cations and anions, penalizing positive
particles that are closest from a distance $s$ to the surfaces. It
mimics, for example, the effect of different sizes for the microions. 
Smaller negative ions are allowed to come in direct contact with the neutral 
surfaces, whereas the positive particles, by their larger size, are held apart from 
an effective distance $s$, related to their sizes. The effect of this term on the system is 
to yield an excess of anions in the region surrounding the 
plates, leading to an inhomogeneity of the local densities
of microions in the vicinity of the uncharged plates. Thus, although 
the surfaces are themselves neutral, this imbalance of microions 
gives rise to a non-vanishing electric field. To allow analytical 
calculations, we shall hereafter assume that $w_s$ has the step-function 
form,
\be
 w_s(\xi)=
\renewcommand{\arraystretch}{1.4}
\left\{\begin{array}{lr}
0, &|\xi|\geq s, \\
\displaystyle\frac1{2s}, &|\xi|< s.
\end{array}\right.
\renewcommand{\arraystretch}{1}
\ee
In the limit $s\to 0$, the function $ w_s(\xi)$ corresponds to the Dirac delta function,
$\displaystyle\delta(\xi)=\lim_{s\to 0}  w_s(\xi)$.

The second term of (\ref{eqn:totalfreeenergy}), $\fpb$,
represents the reduced \textit{bulk excess} 
PB free-energy functional (per unit area),
\bea
\fpb&=& \int\limits_{-\infty}^{\infty} 
\dd x \left\{n_+(x)\left(\ln \left[\Lambda^3 n_+(x)\right]-1\right)+ n_-(x)\left(\ln
\left[\Lambda^3 n_-(x)\right]-1\right)\right.\nonumber\\ 
&&\left.\vphantom{\frac12}\qquad\qquad\qquad\qquad
+\frac12 \phi(x)\left[n_+(x)-n_-(x)\right] 
-\beta\mu \left[n_+(x)+n_-(x)\right] + \beta \Pi_0 \right\} \nonumber\\
&=& \int\limits_{-\infty}^{\infty} 
\dd x \left\{n_+(x)\ln \left[n_+(x)/n_0\right]+ n_-(x)\ln \left[n_-(x)/n_0\right]+
\frac12 \phi(x)\left[n_+(x)-n_-(x)\right]
\right.\nonumber\\ &&\left.\vphantom{\frac12}\qquad\qquad
 -\left[n_+(x)+n_-(x)-2n_0\right]  \right\} ,
\eea
where $\Lambda$ is an arbitrary length scale.
The electrochemical potential and the reference pressure were 
set, respectively, to $\beta\mu=\ln \left(\Lambda^3 n_0\right)$ 
and  $\beta \Pi_0=2n_0$, 
since the system is in electrochemical
equilibrium with the infinite salt reservoir. 
The reduced electrostatic potential, $\phi(x)=\beta e \psi(x)$,
where $e$ is the proton charge and $\psi(x)$ is the electrostatic 
potential, satisfies the Poisson equation,
\be
\frac{\dd^2\phi(x)}{\dd x^2}=
-4\pi\ell\left[n_+(x)-n_-(x)\right] , \label{eqn:poisson}
\ee
where $\ell=\beta e^2/D$ is the Bjerrum length and 
the solvent is treated as a continuum of dielectric constant $D$.

Minimization of the reduced total free-energy functional 
$\bar{f}[n_+(x),n_-(x)]$ with respect to the number densities,
\bea
\frac{\delta\bar{f}[n_+(x),n_-(x)]}{\delta n_+(x)} &=&
\ln \left[n_+(x)/n_0\right]  + \phi(x) + \epsilon \left[
 w_s\left(x+\frac{h}2\right)
+ w_s\left(x-\frac{h}2\right)\right] =0, \\
\frac{\delta\bar{f}[n_+(x),n_-(x)]}{\delta n_-(x)} &=& 
\ln \left[n_-(x)/n_0\right]  - \phi(x)  =0, 
\eea
leads to the Boltzmann distribution for the optimum microion profiles,
\bea
n_+(x)&=& n_0 \exp\left[-\phi(x)-\epsilon w_s\left(x+\frac{h}2\right)
-\epsilon w_s\left(x-\frac{h}2\right) \right],  
\label{eqn:nplus} \\
n_-(x)&=& n_0 \exp\left[\phi(x)\right]. \label{eqn:nminus}
\eea
Replacing~(\ref{eqn:nplus})~and~(\ref{eqn:nminus}) into the Poisson 
equation (\ref{eqn:poisson}) leads to a generalized
PB equation,
\be
\frac{\dd^2\phi(x)}{\dd x^2}=\frac{\kappa^2}2 \left\{ 
\exp\left[\phi(x)\right]
- \exp\left[-\phi(x) - \epsilon w_s\left(x+\frac{h}2\right)
-\epsilon w_s\left(x-\frac{h}2\right) \right]
\right\} ,  \label{eqn:gpb}
\ee
where $\kappa\equiv \sqrt{8\pi n_0 \ell}$ is the inverse
of the Debye screening length.

The appropriate boundary conditions are the vanishing of the 
electrostatic potential and the 
electric field at infinity,
\be
\phi(x\to\pm\infty)=\phi'(x\to\pm\infty)=0; \label{eqn:bound1}
\ee
the vanishing of the electric field at the midplane $(x=0)$,
\be
\phi'(x=0)=0; \label{eqn:bound2}
\ee
the continuity of the electrostatic potential and the electric field across the 
planes located at $x=\pm\frac{h}2\pm s$,
\bea
\phi(x\uparrow \pm\frac{h}2\pm s)&=&\phi(x\downarrow\pm\frac{h}2\pm s), 
\label{eqn:bound3} \\
\phi'(x\uparrow\pm\frac{h}2\pm s)&=&\phi'(x\downarrow\pm\frac{h}2\pm s),
\label{eqn:bound4} 
\eea
where $\displaystyle\phi(x\uparrow y)\equiv \lim_{x\to y_+} \phi(x)$ and  
$\displaystyle\phi(x\downarrow y)\equiv \lim_{x\to y_-} \phi(x)$.  
The boundary conditions (\ref{eqn:bound3}) and (\ref{eqn:bound4}) are based on the fact 
that the charge distribution, which appears on the right-hand side of the 
Poisson equation (\ref{eqn:poisson}), contains just a \textit{finite jump} at 
the planes $x=\pm\frac{h}2\pm s$. 

By symmetry we have $\phi(x)=\phi(-x)$ and we need only to consider 
the \textit{positive} $x$ axis. Because of the non-electrostatic 
depletion of \textit{cations} around the surfaces located at 
$x=\pm\frac{h}2$, the electrostatic potential $\phi(x)$
is \textit{always negative,} since there is an effective excess of 
\textit{anions} around the surfaces.  We shall consider two regimes separately,
namely, the non-overlapping regime $(h>2s)$ and the overlapping regime $(h<2s)$.

\section{The non-overlapping regime, \boldmath$h>2s$}

In the non-overlapping regime, which occurs 
when the separation between the surfaces is larger than the
range of the depletion forces, $h>2s$, the depletion zones 
associated with the two interfaces \textit{do not overlap,} 
and the generalized PB 
equation reads
\be
\frac{\dd^2\phi(x)}{\dd x^2}=
\renewcommand{\arraystretch}{1.4}
\left\{\begin{array}{cc}
\kappa^2 \sinh \phi(x),& 
\mbox{ for }0\leq x\leq\frac{h}2-s \mbox{ and } x\geq\frac{h}2+s,\\
\e^{-2\alpha} \kappa^2\sinh \left[\phi(x)+2\alpha\right], &
\mbox{ for }\frac{h}2-s<x<\frac{h}2+s,
\end{array}\right.
\renewcommand{\arraystretch}{1} \label{eqn:nover_pb}
\ee
where we introduced the parameter $\alpha\equiv\epsilon/8s$. 

Using the identity $\frac{\dd^2\phi(x)}{\dd x^2}=\frac12
\frac{\dd\left[\phi'\right]^2}{\dd\phi}$, 
the non-linear second-order differential equation represented 
by (\ref{eqn:nover_pb}) can be analytically integrated.
Introducing the midplane electrostatic potential, $\fm=\phi(x=0)$, 
and the internal and external electrostatic potentials in the vicinity of the
interface at $x=\frac{h}2$, $\fl=\phi(x=\frac{h}2-s)$ and $\fg=\phi(x=\frac{h}2+s)$,
the solution which satisfy the boundary conditions (\ref{eqn:bound1}) and 
(\ref{eqn:bound2}) can be written explicitly as
\bea
\phi'(x)&=&
\renewcommand{\arraystretch}{1.4}
\left\{\begin{array}{cc}
\kappa \Delta\left[\fm,\phi(x)\right],& \quad\mbox{ for }0\leq 
x\leq\frac{h}2- s, \\
\kappa\sign\left(x-\xii\right)
\e^{-\alpha}\Delta_{\alpha}\left[\fii,\phi(x)\right],&
\quad\mbox{ for }\frac{h}2- s < x <\frac{h}2+s,\\
-2\kappa\sinh\frac{\phi(x)}{2},& \quad\mbox{ for }x\geq\frac{h}2+s,
\end{array}\right.
\renewcommand{\arraystretch}{1} \label{eqn:phiprimex} \\
\phi(x)&=&
\renewcommand{\arraystretch}{1.4}
\left\{\begin{array}{cc}
2\arcsinh \left[\frac{\sinh\frac{\fm}{2}}{\cn\left(\kappa x
\cosh\frac{\fm}{2}, 1/\cosh\frac{\fm}{2}\right)}\right],& \mbox{ for }0\leq 
x\leq\frac{h}2- s, \\
2\arcsinh\left\{\frac{\sinh\left(\frac{\fii}{2}+\alpha\right)}
{\cn\left[\e^{-\alpha}\kappa\left(|x|-\xii\right)\cosh\left(\frac{\fii}{2}+\alpha\right), 
1/\cosh\left(\frac{\fii}{2}+\alpha\right)\right]}\right\}-2\alpha,&
\mbox{ for }\frac{h}2- s < x <\frac{h}2+s,\quad\\
4\arctanh\left\{
\exp\left[-\kappa\left(|x|-\frac{h}2-s\right)\right]
\tanh\frac{\fg}{4}\right\},& \mbox{ for }x\geq\frac{h}2+s,
\end{array}\right.
\renewcommand{\arraystretch}{1} \label{eqn:phix}
\eea
where we introduced
\bea
\Delta(\fm,\phi)&=& 
-{\sqrt{2\cosh\phi-2\cosh\fm}} =
2\sinh\frac{\phi}{2}\,\sqrt{1-\left[{\sinh\frac{\fm}{
2}}/\sinh\frac{\phi}{2}\right]^2},\\
\Delta_{\alpha}(\fii,\phi)&=&
{\sqrt{2\cosh(\phi+2\alpha)-2\cosh(\fii+2\alpha)}} ,
\eea
$\cn(u,k)$ is the Jacobi cosine-amplitude elliptic function with
modulus $k$\cite{gradshteyn,byrd}, $\xii$ is the inversion point where the electric field
vanishes, $\phi'(\xii)=0$, and the electrostatic potential 
$\fii=\phi(\xii)$ is an integration constant to be determined by 
the boundary conditions (\ref{eqn:bound3}) and (\ref{eqn:bound4}).

Matching the electrostatic potential $\phi(x)$ at the planes $x=\frac{h}2\pm s$
by imposing the boundary conditions (\ref{eqn:bound3}) gives  
\bea
\fl&=&2\arcsinh\left\{\frac{\sinh\frac{\fm}{2}}{\cn\left[\kappa\left(\frac{h}2-s\right)
\cosh\frac{\fm}{2}, 1/\cosh\frac{\fm}{2}\right]} \right\}, \label{eqn:phi<} \\
\kappa s &=& 
\frac{{\cal F}_{\alpha}(\fii,\fl) +
{\cal F}_{\alpha}(\fii,\fg)}
{2\e^{-\alpha} \cosh\left(\frac{\fii}{2} +\alpha\right)}, \label{eqn:kappas}\\
\kappa\xii&=&
\frac{{\cal F}\left(\fm,\fl\right)}{\cosh \frac{\fm}{2}}+
\frac{{\cal F}_{\alpha}(\fii,\fl)}
{\e^{-\alpha}\cosh\left(\frac{\fii}{2} +\alpha\right)},
\eea
where we introduced 
\bea
{\cal F}(\fm,\phi)&=&F\left(\arccos\left[{\sinh\frac{\fm}{2}}/
\sinh\frac{\phi}{2}\right],1/\cosh \frac{\fm}{2}\right),\\
{\cal F}_{\alpha}(\fii,\phi)&=& {\cal F}(\fii+2\alpha,\phi+2\alpha)
\nonumber\\&=&
F\left\{\arccos\left[\sinh\left(\frac{\fii}{2}+\alpha\right)/
\sinh\left(\frac{\phi}{2}+\alpha\right)\right],1/\cosh\left(\frac{\fii}{2}+\alpha\right)
\right\} ,
\eea
and $F(\psi,k)=\int\limits_0^\psi {\dd\theta}/{\sqrt{1-k^2 \sin^2\theta}}$ is the 
elliptic integral of the first kind\cite{gradshteyn,byrd}. 

\begin{figure}[hp]
\begin{center}
\leavevmode
\epsfxsize=0.65\textwidth
\epsfbox[15 30 580 430]{"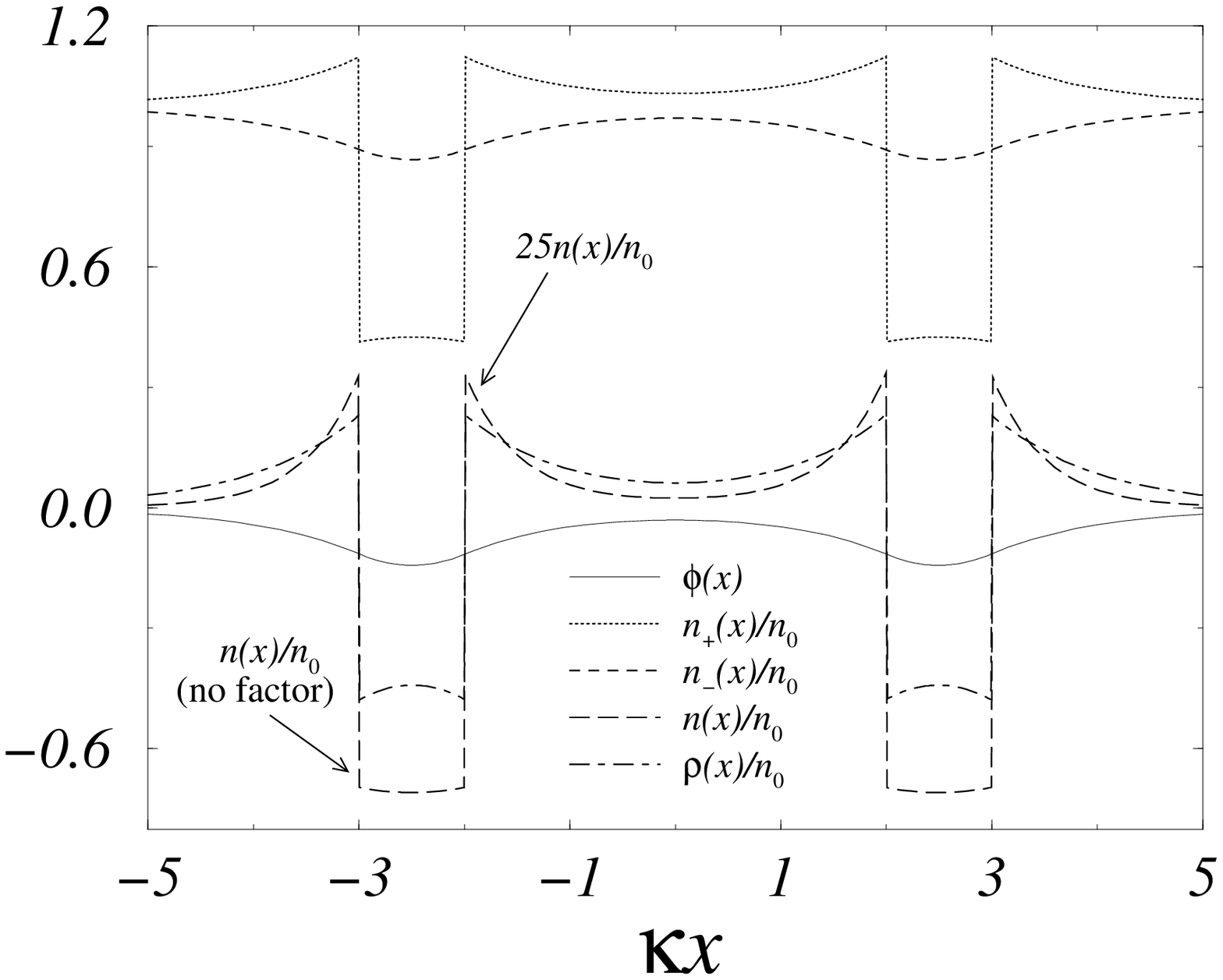"}
\end{center}
\caption{Reduced electrostatic potential $\phi(x)$ and 
density profiles as a function of the distance $x$ for 
the set of parameters $\kappa\epsilon=1$, $\kappa s=1/2$ and $\kappa h=5$ 
(non-overlapping regime). All densities are normalized
to the salt reservoir density, $n_0$.
The positive portion of the particle-density excess $n(x)$ 
was amplified by a factor of 25. 
\label{figure:fig1}}
\begin{center}
\leavevmode
\epsfxsize=0.65\textwidth
\epsfbox[15 30 580 430]{"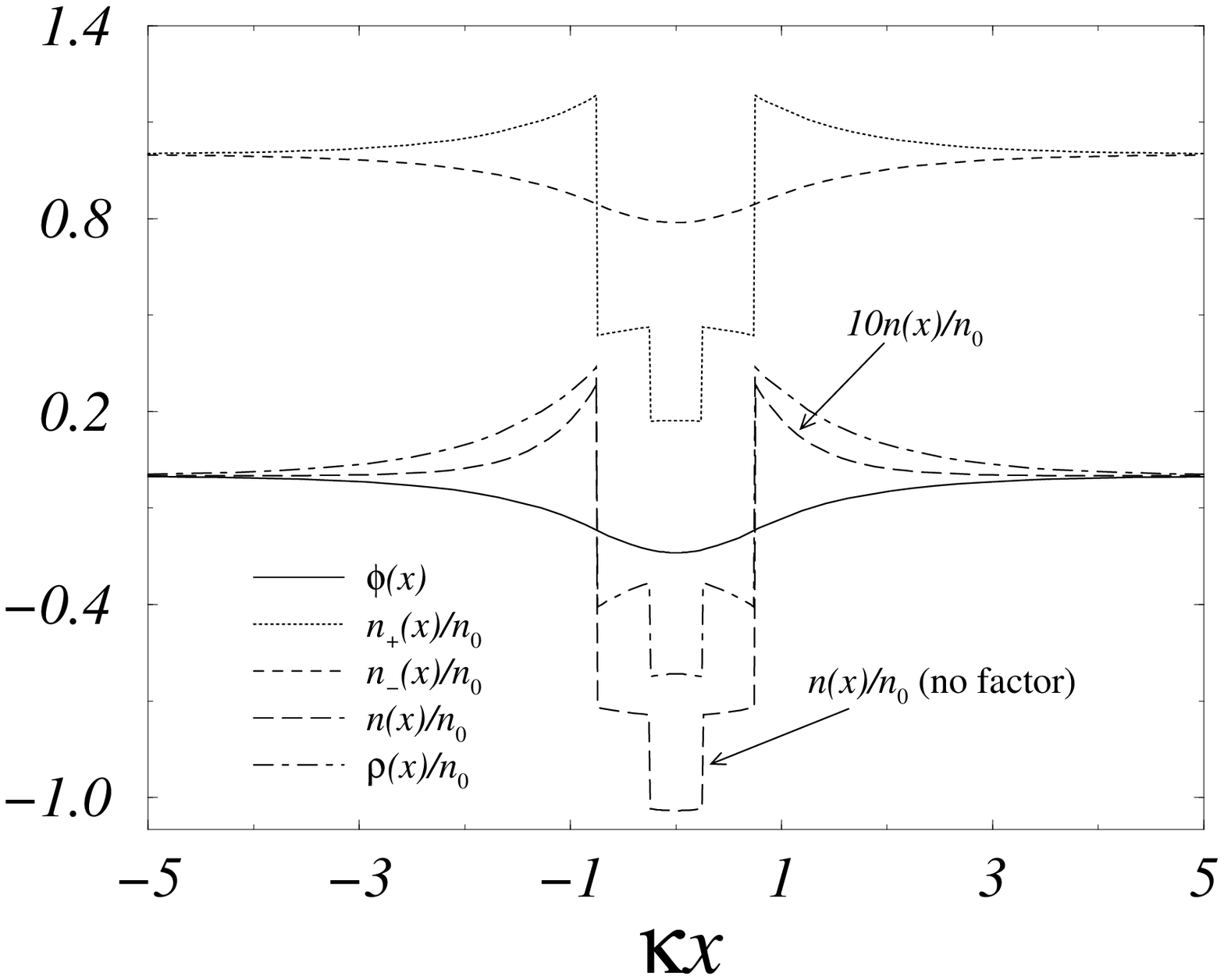"}
\end{center}
\caption{Reduced electrostatic potential $\phi(x)$ and 
density profiles as a function of the distance $x$ for 
the set of parameters
$\kappa\epsilon=1$, $\kappa s=1/2$ and $\kappa h=1/2$ 
(overlapping regime). All number densities are normalized
to the salt reservoir density, $n_0$. The positive portion of 
the particle-density excess $n(x)$ was amplified 
by a factor of 10. 
\label{figure:fig2}}
\end{figure}

On the other hand, matching the electric field $\phi'(x)$ at the planes $x=\frac{h}2\pm s$
by imposing the boundary conditions (\ref{eqn:bound4}) leads to 
\bea
\e^{-2\alpha} \left[\cosh(\fl+2\alpha)-\cosh(\fg+2\alpha)\right]&=&
\cosh\fl - \cosh\fm -2\sinh^2\frac{\fg}{2},\qquad\label{eqn:final}\\
\cosh(\fii+2\alpha)&=&\cosh(\fg+2\alpha)-2\e^{2\alpha}\sinh^2\frac{\fg}{2} .  
\label{eqn:ft}
\eea
Equations~(\ref{eqn:kappas})~and~(\ref{eqn:final}) represent a pair 
of coupled equations which can be solved for $\fm$ and $\fg$, 
since we can use (\ref{eqn:phi<}) and (\ref{eqn:ft}) to eliminate 
$\fl$ and $\fii$, respectively. Once obtained $\fm$ and $\fg$ 
which solve equations~(\ref{eqn:kappas})~and~(\ref{eqn:final}), 
the electrostatic potential $\phi(x)$
can be obtained by replacing them into the closed 
expression~(\ref{eqn:phix}). To illustrate typical profiles for the non-overlapping 
regime, in Figure~(\ref{figure:fig1}) we show the reduced electrostatic potential 
$\phi(x)$ and the density profiles $n_\pm(x)$ for a fixed value of 
$\epsilon,s$ and $h$. We also present the particle-density excess over the 
reservoir,
\be
n(x)\equiv n_+(x)+n_-(x)-2n_0,
\ee
and the charge density, 
\be
\rho(x)\equiv n_+(x)-n_-(x).
\ee

\begin{figure}[hp]
\begin{center}
\leavevmode
\epsfxsize=0.65\textwidth
\epsfbox[15 30 580 430]{"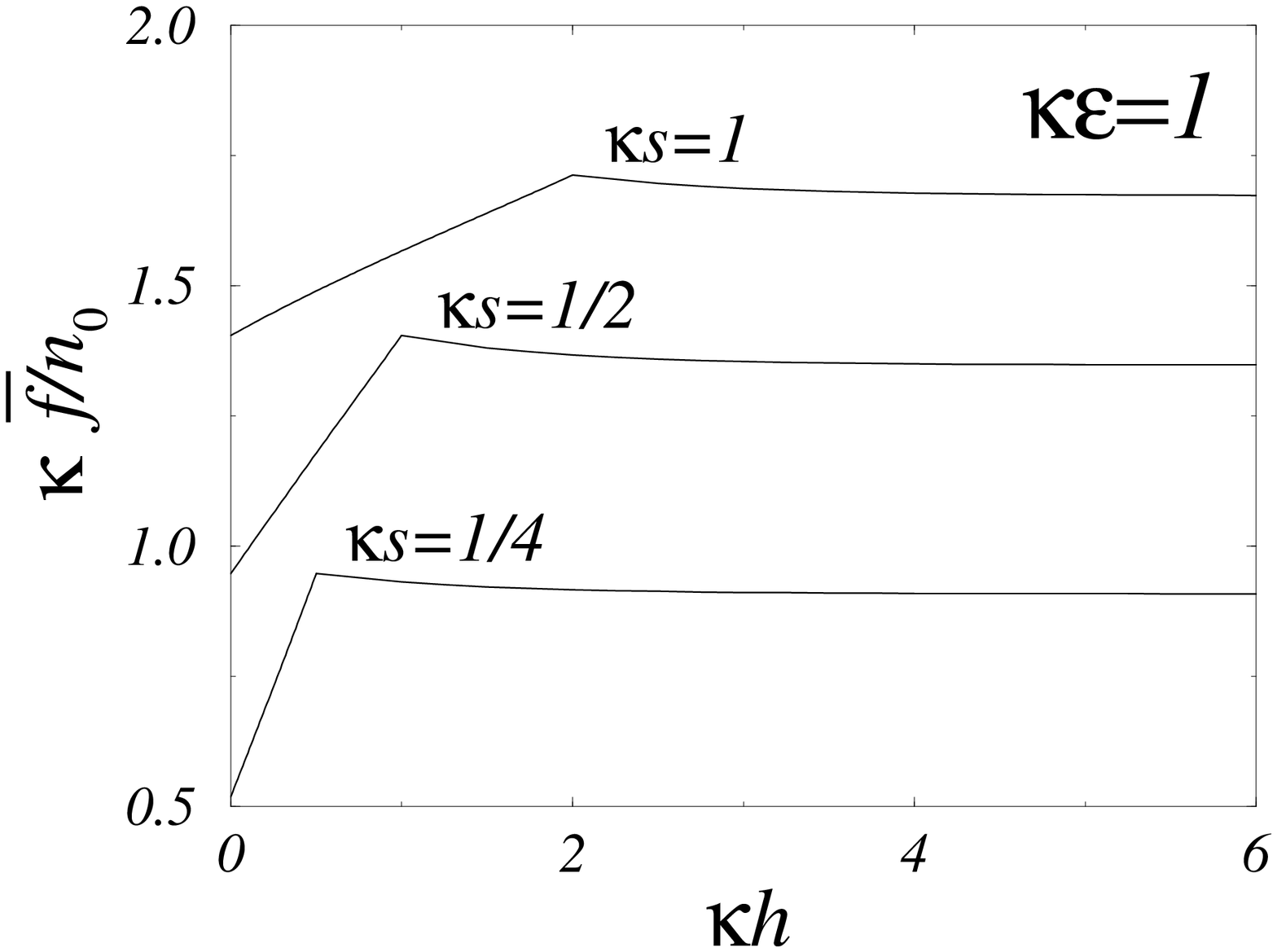"}
\end{center}
\caption{Reduced total free-energy density $\bar{f}$ as
a function of the separation of the surfaces $h$ for 
a fixed value of the depletion strength $(\kappa\epsilon=1)$
and three values of the depletion range $(\kappa s=1/4, 1/2, 1)$.
Although the free-energy density itself is continuous upon crossing the 
separation $h=2s$, it has a kink at this special value, giving rise
to a change between attractive and repulsive forces (see next graph). 
\label{figure:fig3}}
\begin{center}
\leavevmode
\epsfxsize=0.65\textwidth
\epsfbox[15 30 580 430]{"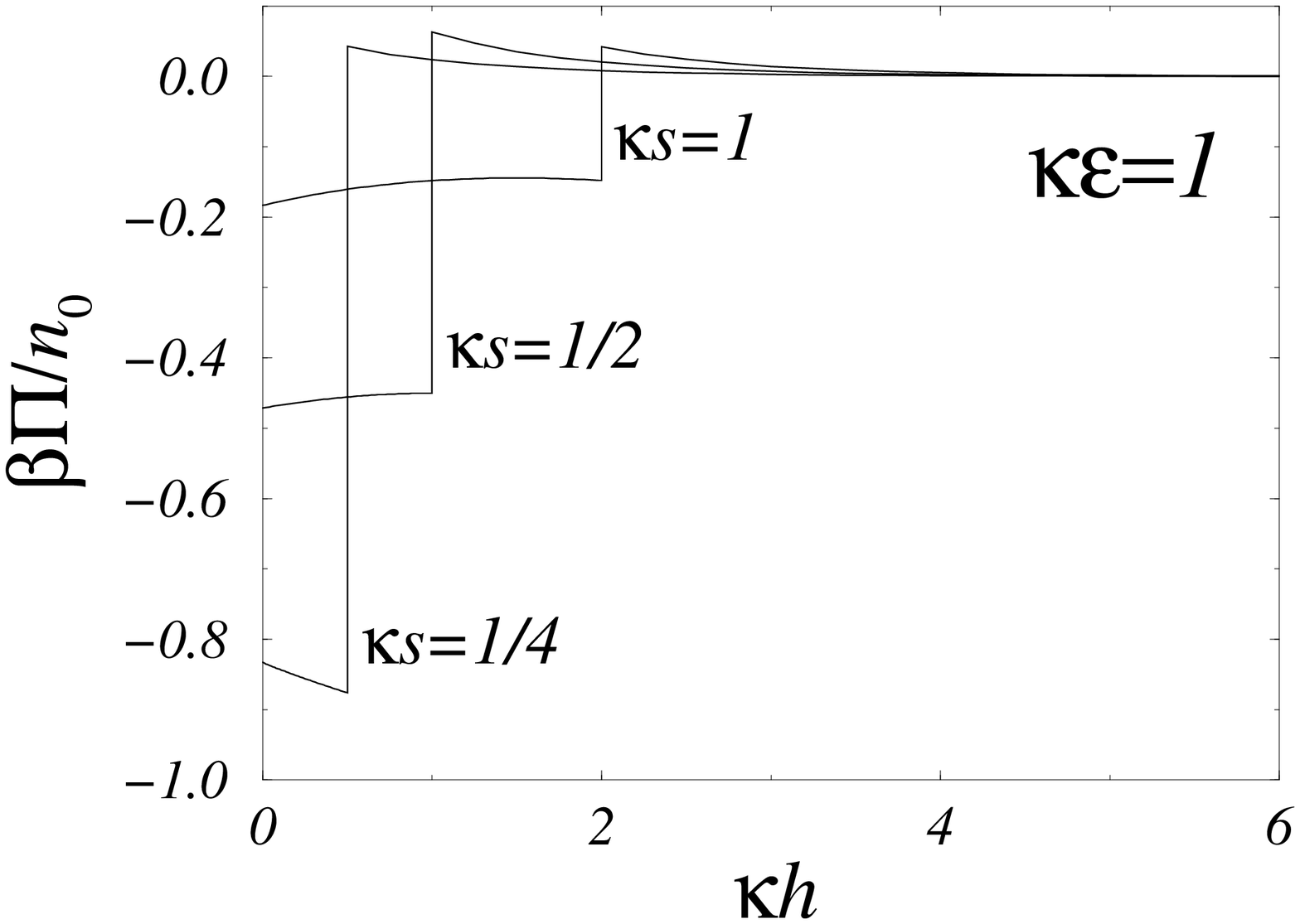"}
\end{center}
\caption{Disjoining pressure $\Pi$ as a function of the 
separation of the surfaces $h$ for a fixed value of the 
depletion strength $(\kappa\epsilon=1)$ and three values of 
the depletion range $(\kappa s=1/4, 1/2, 1)$. This graph corresponds to 
the negative derivative of the curves of the previous figure. 
Note the discontinuity of the pressure at the separation $h=2s$, 
associated with the kink of the free-energy.
\label{figure:fig4}}
\end{figure}

The total free-energy density associated with the electrostatic 
potential (\ref{eqn:phix}) and the microion profiles 
(\ref{eqn:nplus}) and (\ref{eqn:nminus}) is obtained by replacing 
their closed forms into the total free-energy functional, given 
by (\ref{eqn:totalfreeenergy}), and performing the integrations.
After some algebra, we obtain 
\bea
\frac{\kappa}{n_0} \bar{f}&=&
8\kappa s \left(1-\e^{-2\alpha}\right)+
2\Delta(\fm,\fl)\left(\fl-4 \coth\frac{\fl}2 \right)
+16 {\cal E}(\fm,\fl) \cosh\frac{\fm}2 \nonumber\\
&&-8 {\cal F}(\fm,\fl)\,\frac{\sinh^2\frac{\fm}2}{\cosh\frac{\fm}2}
+2\e^{-\alpha} {\Delta_{\alpha}(\fii,\fl)}\left[\fl -4\coth\left(\frac{\fl}{2}
+\alpha\right) \right]\nonumber\\
&&+2\e^{-\alpha}{\Delta_{\alpha}(\fii,\fg)}
 \left[\fg -4\coth\left(\frac{\fg}{2}+\alpha\right) \right]
+16\e^{-\alpha}\left[{\cal E}_{\alpha}(\fii,\fl)+{\cal E}_{\alpha}(\fii,\fg)\right] 
\cosh\left(\frac{\fii}2+\alpha\right)\!\!\! \nonumber\\
&&-8\e^{-\alpha}\left[{\cal F}_{\alpha}(\fii,\fl)+{\cal F}_{\alpha}(\fii,\fg)\right]
\frac{\sinh^2\left(\frac{\fii}2+\alpha\right)} {\cosh\left(\frac{\fii}2+\alpha\right)}
+4\sinh\frac{\fg}2  \left(\fg-4 \tanh\frac{\fg}4 \right) , 
\label{eqn:freenoover}
\eea
where we introduced 
\bea
{\cal E}(\fm,\fl)&=& E\left(\arccos\left[\sinh\frac{\fm}{2}/
\sinh\frac{\fl}{2}\right],1/\cosh \frac{\fm}{2}\right), \label{eqn:elliptic_e} \\
{\cal E}_{\alpha}(\fii,\phi) &=& {\cal E}(\fii+2\alpha,\phi+2\alpha) \nonumber\\
&=& E\left\{\arccos\left[\sinh\left(\frac{\fii}{2}+\alpha\right)/
\sinh\left(\frac{\phi}{2}+\alpha\right)\right],1/\cosh\left(\frac{\fii}{2}+\alpha\right)
\right\}, 
\eea 
and $E(\psi,k)=\int\limits_0^\psi {\dd\theta} {\sqrt{1-k^2 \sin^2\theta}}$ is the 
elliptic integral of the second kind\cite{gradshteyn,byrd}. 
The closed analytical expression (\ref{eqn:freenoover}) was checked against 
numerical integration of the free-energy density (\ref{eqn:totalfreeenergy}).
At the end of Section 4, in Figure~(\ref{figure:fig3}), we present the total 
free-energy density as a function of the separation of the surfaces $h$ for a 
fixed value of the depletion strength $\epsilon$ and several values of the 
depletion range $s$.

The disjoining pressure $\Pi$ is given by the negative derivative 
of the total free energy with respect to the separation of
the surfaces, $h$, for constant depletion strength, $\epsilon$, 
and range, $s$,
\be
\beta\Pi \equiv \left. -\kappa 
\frac{\partial\bar{f}}{\partial\bar{h}}\right|_{\alpha,\bar{s}}, 
\ee
where we introduced the dimensionless
distances, $\bar{h}=\kappa h$ and  $\bar{s}=\kappa s$. 
After a lengthy calculation (see Appendix), we obtain a 
very simple final expression for the disjoining pressure,  
\be
\beta\Pi = 4n_0 \sinh^2 \frac{\fm}2
= n(x=0). 
\label{eqn:pressurenoover}
\ee
The above simple analytical expression was checked against 
numerical differentiation of the free-energy density for the 
non-overlapping regime (\ref{eqn:freenoover}).
Thus, it turns out that the disjoining pressure for the 
non-overlapping regime is given simply by the 
excess osmotic pressure of the microions at the midplane over 
the bulk (reservoir) pressure. Although it might be tempting to 
attribute this simple result to the contact-value theorem
for charged plates\cite{andelman,israelachvili,safran}, 
we stress that this is not the case. Actually, an expression 
for the particle-density excess similar to the charged-plates 
case, 
\be
n(x)=n(x=0) + \frac{n_0}{\kappa^2} \left[\phi'(x)\right]^2,
\ee
holds only for $0\leq |x| \leq \left| \frac{h}2-s\right|$. 
Since there are non-vanishing discontinuities for the density of cations 
$n_+(x)$ upon crossing the surfaces at $x=\pm\frac{h}2\pm s$,
\bea
\Delta n_+\left(x=\left|\frac{h}2-s\right|\right)&\equiv&
n_+\left(x\uparrow
\left|\frac{h}2-s\right|\right)-n_+\left(x\downarrow\left|\frac{h}2-s\right|\right), 
\label{eqn:nplusdisc} \\
\Delta n_+\left(x=\left|\frac{h}2+s\right|\right)&\equiv&
n_+\left(x\uparrow
\left|\frac{h}2+s\right|\right)-n_+\left(x\downarrow\left|\frac{h}2+s\right|\right),
\eea  
the corrected expressions for the particle-density excess for 
$|x|> \left| \frac{h}2-s\right|$ are
\bea
n(x)&=&n(x=0)  +
\Delta n_+\left(x=\left|\frac{h}2-s\right|\right) 
+ \frac{n_0}{\kappa^2} \left[\phi'(x)\right]^2
\nonumber\\
&=& n(x=\xii) + \frac{n_0}{\kappa^2} \left[\phi'(x)\right]^2 , 
\qquad\mbox{ for } \left|\frac{h}2-s\right|<x<
\left|\frac{h}2+s\right|, \\
n(x)&=&n(x=0)+\Delta n_+\left(x=\left|\frac{h}2-s\right|\right)
+\Delta n_+\left(x=\left|\frac{h}2+s\right|\right) 
+\frac{n_0}{\kappa^2} \left[\phi'(x)\right]^2 \nonumber\\
&=&\frac{n_0}{\kappa^2} \left[\phi'(x)\right]^2, 
\qquad\mbox{ for } x>\left|\frac{h}2+s\right|,
\eea
where we used (\ref{eqn:final}) and (\ref{eqn:ft}) to simplify
the above expressions. However, these additional contributions 
cancel when we evaluate the 
disjoining pressure (for the non-overlapping regime),
and we obtain (\ref{eqn:pressurenoover}), a result similar to the
contact-value theorem expression for charged plates. 
Since the disjoining pressure (\ref{eqn:pressurenoover}) is always 
positive, the interaction between the surfaces
for the non-overlapping regime is \textit{always repulsive.}
At the end of Section 4, in Figure~(\ref{figure:fig4}), we present the 
disjoining pressure as a function of the separation of the surfaces $h$ for a 
fixed value of the depletion strength $\epsilon$ and several values of the 
depletion range $s$.

\section{The overlapping regime, \boldmath$h<2s$}

In the overlapping regime, which occurs 
when the separation between the surfaces is smaller than the range 
of the depletion forces, $h<2s$, the depletion zones associated with 
the two interfaces \textit{do overlap,} and the generalized PB 
equation reads
\be
\frac{\dd^2\phi(x)}{\dd x^2}=
\renewcommand{\arraystretch}{1.4}
\left\{\begin{array}{cc}
\e^{-4\alpha} \kappa^2\sinh \left[\phi(x)+4\alpha\right],& 
\mbox{ for }0\leq x\leq s-\frac{h}2,\\
\e^{-2\alpha} \kappa^2\sinh \left[\phi(x)+2\alpha\right], &
\quad \mbox{ for }s-\frac{h}2<x<s+\frac{h}2, \\
\kappa^2 \sinh \phi(x),& 
\mbox{ for }x\geq s+\frac{h}2 .
\end{array}\right.
\renewcommand{\arraystretch}{1}
\ee

The calculation is analogous to the case when there is no overlapping
of the depletion zones, $h>2s$. Now the pair of coupled equations to be 
solved for $\fm=\phi(x=0)$ and $\fg=\phi(x=s-\frac{h}2 )$ is given by 
\be
\kappa s = 
\frac{{\cal F}_{2\alpha}\left(\fm,\fl\right)}{\e^{-2\alpha}
\cosh\left(\frac{\fm}{2}+2\alpha\right)}+
\frac{{\cal F}_{\alpha}(\fii,\fg) -
{\cal F}_{\alpha}(\fii,\fl)}
{2\e^{-\alpha} \cosh\left(\frac{\fii}{2} +\alpha\right)},
\label{eqn:over1}
\ee
and 
\be
\e^{-2\alpha} \left[\cosh(\fl+2\alpha)-\cosh(\fg+2\alpha)\right]=
\e^{-4\alpha} \left[\cosh(\fl+4\alpha)-\cosh(\fm+4\alpha)\right]
-2\sinh^2\frac{\fg}{2}, \label{eqn:over2}
\ee
where $\fl=\phi(x=s-\frac{h}2)$ and $\fii=\phi(x=\xii)$ are eliminated by using the relations
\bea
\fl&=&2\arcsinh\left\{\frac{\sinh\left(\frac{\fm}{2}+2\alpha\right)}
{\cn\left[\e^{-2\alpha}\kappa\left(s-\frac{h}2\right)
\cosh\left(\frac{\fm}{2}+2\alpha\right), 
1/\cosh\left(\frac{\fm}{2}+2\alpha\right)\right]} \right\} - 4\alpha, \qquad\ \\
\cosh(\fii+2\alpha)&=&\cosh(\fg+2\alpha)-2\e^{2\alpha}\sinh^2\frac{\fg}{2} .
\eea
Once solved the system (\ref{eqn:over1}) and (\ref{eqn:over2}), 
the electric field and the electrostatic potential can be obtained by 
replacing the solution $(\fm,\fg)$ into the closed expressions, 
\bea
\phi'(x)&=&
\renewcommand{\arraystretch}{1.4}
\left\{\begin{array}{cc}
\kappa\e^{-2\alpha}\Delta_{2\alpha}\left[\fm,\phi(x)\right],&
\quad\mbox{ for }0\leq 
x\leq s-\frac{h}2, \\
\kappa\sign\left(x-\xii\right)
\e^{-\alpha}\Delta_{\alpha}\left[\fii,\phi(x)\right],&
\quad\mbox{ for }s-\frac{h}2 < x <s+\frac{h}2,\\
-2\kappa\sinh\frac{\phi(x)}{2},& \quad\mbox{ for }x\geq s+\frac{h}2 , 
\end{array}\right.
\renewcommand{\arraystretch}{1} \\
\phi(x)&=&
\renewcommand{\arraystretch}{1.4}
\left\{\begin{array}{cc}
2\arcsinh\left\{\frac{\sinh\left(\frac{\fm}{2}+2\alpha\right)}
{\cn\left[\e^{-2\alpha}\kappa x\cosh\left(\frac{\fm}{2}+2\alpha\right), 
1/\cosh\left(\frac{\fm}{2}+2\alpha\right)\right]}\right\}-4\alpha, &
\mbox{ for } 0\leq x\leq s-\frac{h}2,  \\
2\arcsinh\left\{\frac{\sinh\left(\frac{\fii}{2}+\alpha\right)}
{\cn\left[\e^{-\alpha}\kappa (|x|-\xii)\cosh\left(\frac{\fii}{2}+\alpha\right), 
1/\cosh\left(\frac{\fii}{2}+\alpha\right)\right]}\right\}-2\alpha,
&\mbox{ for } s-\frac{h}2<x<s+\frac{h}2, \quad\\
4\arctanh\left\{
\exp\left[-\kappa\left(|x|-\frac{h}2-s\right)\right]
\tanh\frac{\fg}{4}\right\},& \mbox{ for } x\geq s+\frac{h}2, 
\end{array}\right.
\renewcommand{\arraystretch}{1}
\eea
with the inversion point given by 
\be
\kappa \xii = 
\frac{{\cal F}_{2\alpha}\left(\fm,\fl\right)}{\e^{-2\alpha}
\cosh\left(\frac{\fm}{2}+2\alpha\right)}-
\frac{{\cal F}_{\alpha}(\fii,\fl)}
{\e^{-\alpha}\cosh\left(\frac{\fii}{2} +\alpha\right)} . 
\ee
To illustrate typical profiles for the overlapping 
regime, in Figure~(\ref{figure:fig2}) we show the reduced electrostatic 
potential $\phi(x)$ and the density profiles $n_{\pm}(x), n(x)$ and 
$\rho(x)$ for a fixed value of $\epsilon,s$ and $h$.

Again, it is possible to obtain a closed analytical expression
for the total free-energy density,
\bea
\frac{\kappa}{n_0} \bar{f}&=& 4\left(\bar{s}-\frac{\bar{h}}{2}\right)
\left(1-\e^{-4\alpha}\right)+
4 \bar{h}\left(1-\e^{-2\alpha}\right)+\frac{\kappa}{n_0}\bar{f}_{\mathrm{aux}}, 
\label{eqn:freeover} \\
\frac{\kappa}{n_0}\bar{f}_{\mathrm{aux}} &=& 
2\e^{-2\alpha}{\Delta_{2\alpha}(\fm,\fl)}\left[\fl -4\coth\left(\frac{\fl}{2}
+2\alpha\right) \right]\nonumber\\
&& +16\e^{-2\alpha}{\cal E}_{2\alpha}(\fm,\fl)\cosh\left(\frac{\fm}2+2\alpha\right)
-8\e^{-2\alpha}{\cal F}_{2\alpha}(\fm,\fl)\,\frac{\sinh^2\left(\frac{\fm}2+2\alpha\right)}
{\cosh\left(\frac{\fm}2+2\alpha\right)} \nonumber\\
&&-2\e^{-\alpha}{\Delta_{\alpha}(\fii,\fl)}
 \left[\fl -4\coth\left(\frac{\fl}{2}+\alpha\right) \right] +2\e^{-\alpha}
{\Delta_{\alpha}(\fii,\fg)}\left[\fg -4\coth\left(\frac{\fg}{2}
+\alpha\right) \right]\nonumber\\
&&-16\e^{-\alpha}\left[{\cal E}_{\alpha}(\fii,\fl)-{\cal E}_{\alpha}(\fii,\fg)\right] 
\cosh\left(\frac{\fii}2+\alpha\right) \nonumber\\
&&+8\e^{-\alpha}\left[{\cal F}_{\alpha}(\fii,\fl)-{\cal F}_{\alpha}(\fii,\fg)\right]
\frac{\sinh^2\left(\frac{\fii}2+\alpha\right)} {\cosh\left(\frac{\fii}2+\alpha\right)}
+4\sinh\frac{\fg}2  \left(\fg-4 \tanh\frac{\fg}4 \right) , 
\label{eqn:freeauxover}
\eea
which leads, after some algebra (see Appendix), 
to a simple expression for the disjoining pressure,
\bea
\beta\Pi \equiv \left. -\kappa 
\frac{\partial\bar{f}}{\partial\bar{h}}\right|_{\alpha,\bar{s}} 
&=& 2n_0 \left[\e^{-4\alpha}\cosh\left(\fm+4\alpha \right)-1+2\e^{-\fl-6\alpha}
\sinh 2\alpha\right] \nonumber\\
&=& n(x=0)+ 2\Delta n_+(x=s-\frac{h}2)  .
\eea
The above simple analytical expression was checked against 
numerical differentiation of the free-energy density for the 
overlapping regime (\ref{eqn:freeover}).
It should be remarked that, in this case, 
the disjoining pressure \textit{has not the form} of the expression
given by the contact-value theorem for charged
plates\cite{andelman,israelachvili,safran}. 
An additional contribution due to the discontinuity 
$\Delta n_+(x=s-\frac{h}2)$ of the density of cations upon 
crossing the surface located at $x=s-\frac{h}2$, appears.
Contrary to the non-overlapping regime, this additional 
contribution does not cancel when we evaluate the disjoining
pressure. According to this imbalanced pressure acting onto 
the neutral surfaces, this leads to an effective attraction 
between them.

Figures~(\ref{figure:fig3})~and~(\ref{figure:fig4})
show the total free-energy density and the 
associated pressure as a function of the separation of the 
surfaces $h$ for a fixed value of the depletion strength $\epsilon$
and several values of the depletion range $s$.
The nature of the interactions changes from attractive to 
repulsive at a separation $h=2s$. 

\section{Concluding remarks}

We have proposed a new mechanism for attraction
between neutral plates immersed in a monovalent electrolyte 
solution, which does not include any correlation or fluctuation
effects. The electrostatic potential and the density profiles of 
the microions are obtained from analytical solutions of the generalized
PB equations, which include non-electrostatic depletion interactions. 
Explicit analytical expressions of all thermodynamical properties, 
including the disjoining pressure, were obtained.  

We found that the repulsive interactions at large separations 
become attractive when the separation between the plates is decreased. 
The range of the attractive forces is closely 
related to the range of the non-electrostatic depletion
interactions introduced in the formulation of the model. Although 
this result is not at all surprising, since the attraction is
induced by the imbalanced pressure originated from the ionic depletion
in the region between the two approaching surfaces, 
we found that the disjoining pressure \textit{has not the form}
of the expression given by the contact-value theorem for charged plates.

The proposed mechanism could mimic neutral surfaces immersed in
an electrolyte solution containing ions of different sizes. We
expect to observe attraction when the separation between the 
two surfaces is comparable to the size of the smaller ions.
We stress the fact that we do not include any non-mean-field 
effects to obtain attractive forces. Surely, for short 
separations, other features should be taken into account, 
as for example, the discreteness of the charges and ionic 
correlations. However, using an exactly solvable model, 
we showed that the inclusion of non-mean-field  
effects is not a \textit{necessary condition} to obtain attractive 
interactions. 

\appendix
\section{The disjoining pressure}

The disjoining pressure $\Pi$ is given by the negative derivative 
of the total free energy with respect to the separation of
the surfaces, $h$, for constant depletion strength and range,
$\epsilon$ and $s$,
\be
\frac{\beta\Pi}{n_0} \equiv \left. -\frac{\kappa}{n_0} 
\frac{\partial\bar{f}}{\partial\bar{h}}\right|_{\alpha,\bar{s}}
= 
-\frac{\kappa}{n_0}\frac{\dd \bar{f}}{\dd \fm}
\left.\frac{\partial\fm}{\partial\bar{h}}\right|_{\alpha,\bar{s}}
-\frac{\kappa}{n_0}\frac{\dd \bar{f}}{\dd \fl}
\left.\frac{\partial\fl}{\partial\bar{h}}\right|_{\alpha,\bar{s}}
-\frac{\kappa}{n_0}\frac{\dd \bar{f}}{\dd \fii}
\left.\frac{\partial\fii}{\partial\bar{h}}\right|_{\alpha,\bar{s}}
-\frac{\kappa}{n_0}\frac{\dd \bar{f}}{\dd \fg}
\left.\frac{\partial\fg}{\partial\bar{h}}\right|_{\alpha,\bar{s}} ,
\label{eqn:generalpress}
\ee
where we introduced, for convenience, the dimensionless
distances, $\bar{h}=\kappa h$ and  $\bar{s}=\kappa s$. 
The (four) derivatives of the free energy which appears into 
(\ref{eqn:generalpress}), $\dd\bar{f}/\dd\varphi$,
with $\varphi=(\fm,\fl,\fii,\fg)$, can be obtained directly by 
using the free-energy expressions (\ref{eqn:freenoover}) and 
(\ref{eqn:freeover}). On the other hand, the partial 
derivatives $\partial\varphi/\partial\bar{h}|_{\alpha,\bar{s}}$
are obtained by the matrix product
\be
\left(
\frac{\partial\fm}{\partial\bar{h}},
\frac{\partial\fl}{\partial\bar{h}},
\frac{\partial\fii}{\partial\bar{h}},
\frac{\partial\fg}{\partial\bar{h}}
\right)_{\alpha,\bar{s}} = 
\left[\frac{\partial(\bar{h},\bar{s},u,v)}{\partial(\fm,\fl,\fii,\fg)}\right]^{-1}
(1,0,0,0),
\ee
where $u$ and $v$ are the boundary conditions (\ref{eqn:bound4}) 
written in a parametric form involving $\fm,\fl,\fii$ and $\fg$. 
We will give the explicit expressions of $u$ and $v$ (and their derivatives)
when we treat separately the non-overlapping and the overlapping regimes. 

\subsection{The non-overlapping regime}

For the non-overlapping regime, the total free-energy density
is given by (\ref{eqn:freenoover}), with derivatives 
\bea
\frac{\kappa}{n_0} \frac{\dd\bar{f}}{\dd\fm}&=&
4 {\cal E}(\fm,\fl) \sinh\frac{\fm}2 -\frac{2\sinh\fm}{\Delta(\fm,\fl)}
\left(\fl-2 \tanh^2\frac{\fm}2\,\coth\frac{\fl}2 \right),\\
\frac{\kappa}{n_0} \frac{\dd\bar{f}}{\dd\fl}&=&
\frac{2\sinh\fl}{\Delta(\fm,\fl)}\left(\fl-2\tanh\frac{\fl}2 \right)
+\frac{ 2\e^{-\alpha} \sinh\left(\fl+2\alpha\right)}
{\Delta_\alpha(\fii,\fl)}\left[\fl -2\tanh\left(\frac{\fl}{2}+\alpha\right) 
\right],\quad\ \\
\frac{\kappa}{n_0} \frac{\dd\bar{f}}{\dd\fii}&=&
4\e^{-\alpha}
\left[{\cal E}_{\alpha}(\fii,\fl)+{\cal E}_{\alpha}(\fii,\fg)\right] 
\sinh\left(\frac{\fii}2+\alpha\right) \nonumber\\
&&-\frac{2\e^{-\alpha}\sinh\left(\fii+2\alpha\right)}{\Delta_\alpha(\fii,\fl)}
\left[\fl-2 \tanh^2\left(\frac{\fii}2+\alpha\right)
\coth\left(\frac{\fl}2+\alpha \right)\right] \nonumber\\
&&-\frac{2\e^{-\alpha}\sinh\left(\fii+2\alpha\right)}{\Delta_\alpha(\fii,\fg)}
\left[\fg-2 \tanh^2\left(\frac{\fii}2+\alpha\right)
\coth\left(\frac{\fg}2+\alpha \right)\right] ,  \\
\frac{\kappa}{n_0} \frac{\dd\bar{f}}{\dd\fg}&=&
\frac{2\e^{-\alpha} \sinh\left(\fg+2\alpha\right)}
{\Delta_{\alpha}(\fii,\fg)}\left[\fg -2\tanh\left(\frac{\fg}{2}+\alpha\right)\right] 
+ 2\cosh\frac{\fg}2 \left(\fg-2 \tanh\frac{\fg}2\right) .
\eea

The defining equations for $\bar{h},\bar{s},u$ and $v$ are 
\bea
\bar{h}&=& \frac{2{\cal F}\left(\fm,\fl\right)}{\cosh \frac{\fm}{2}}+
\frac{{\cal F}_{\alpha}(\fii,\fl) +
{\cal F}_{\alpha}(\fii,\fg)}
{\e^{-\alpha} \cosh\left(\frac{\fii}{2} +\alpha\right)},\\
\bar{s}&=& 
\frac{{\cal F}_{\alpha}(\fii,\fl) +
{\cal F}_{\alpha}(\fii,\fg)}
{2\e^{-\alpha} \cosh\left(\frac{\fii}{2} +\alpha\right)}, \\
u&=&\e^{-\alpha}\Delta_\alpha(\fii,\fl)+\Delta(\fm,\fl)=0,\\
v&=&\e^{-\alpha}\Delta_\alpha(\fii,\fg)+2\sinh\frac{\fg}2=0,
\eea
and their derivatives necessary
for the evaluation of the Jacobian
${\partial(\bar{h},\bar{s},u,v)}/{\partial(\fm,\fl,\fii,\fg)}$,
\bea
\frac{\dd\bar{h}}{\dd\fm}&=& 
-\frac{{\cal E}\left(\fm,\fl\right)}{\sinh\frac{\fm}2}
-\frac{2\tanh\frac{\fm}2\,\coth\frac{\fl}2}{\Delta(\fm,\fl)}, \\
\frac{\dd\bar{h}}{\dd\fl}&=& \frac2{\Delta(\fm,\fl)} +
\frac1{\e^{-\alpha}\Delta_\alpha(\fii,\fl)} , \\
\frac{\dd\bar{h}}{\dd\fii}&=&
-\frac{\left[{\cal E}_{\alpha}(\fii,\fl)+{\cal E}_{\alpha}(\fii,\fg)\right]}
{2\e^{-\alpha}\sinh\left(\frac{\fii}2+\alpha\right)}
-\left[ \frac{\coth\left(\frac{\fl}2+\alpha\right)}{\e^{-\alpha}\Delta_\alpha(\fii,\fl)}
+ \frac{\coth\left(\frac{\fg}2+\alpha\right)}
{\e^{-\alpha}\Delta_\alpha(\fii,\fg)}\right] 
\tanh\left(\frac{\fii}2+\alpha\right) , \\
\frac{\dd\bar{h}}{\dd\fg} &=& \frac1{\e^{-\alpha}\Delta_\alpha(\fii,\fg)}, \\
\frac{\dd\bar{s}}{\dd\fm} &=& 0,\\
\frac{\dd\bar{s}}{\dd\fl} &=& \frac1{2\e^{-\alpha}\Delta_\alpha(\fii,\fl)}, \\
\frac{\dd\bar{s}}{\dd\fii} &=& 
-\frac{\left[{\cal E}_{\alpha}(\fii,\fl)+{\cal E}_{\alpha}(\fii,\fg)\right]}
{4\e^{-\alpha}\sinh\left(\frac{\fii}2+\alpha\right)}
-\left[\frac{\coth\left(\frac{\fl}2+\alpha\right)}
{2\e^{-\alpha}\Delta_\alpha(\fii,\fl)}+
\frac{\coth\left(\frac{\fg}2+\alpha\right)}
{2\e^{-\alpha}\Delta_\alpha(\fii,\fg)}\right]
\tanh\left(\frac{\fii}2+\alpha\right) ,\qquad \\ 
\frac{\dd\bar{s}}{\dd\fg} &=& \frac1{2\e^{-\alpha}\Delta_\alpha(\fii,\fg)}, \\
\frac{\dd u}{\dd\fm} &=& -\frac{\sinh\fm}{\Delta(\fm,\fl)}, \\
\frac{\dd u}{\dd\fl} &=& \frac{\sinh\fl}{\Delta(\fm,\fl)} + 
\frac{\e^{-\alpha}\sinh\left(\fl+2\alpha\right)} {\Delta_\alpha(\fii,\fl)}, \\
\frac{\dd u}{\dd\fii} &=& -\frac{\e^{-\alpha}\sinh\left(\fii+2\alpha\right)}
{\Delta_\alpha(\fii,\fl)}, \\
\frac{\dd u}{\dd\fg} &=& \frac{\dd v}{\dd\fm} = \frac{\dd v}{\dd\fl} = 0,\\
\frac{\dd v}{\dd\fii} &=& 
-\frac{\e^{-\alpha}\sinh\left(\fii+2\alpha\right)} {\Delta_\alpha(\fii,\fg)}, \\
\frac{\dd v}{\dd\fg} &=&\frac{\e^{-\alpha}\sinh\left(\fg+2\alpha\right)}{\Delta_\alpha(\fii,\fg)}
+\cosh\frac{\fg}2.
\eea

Putting all together into the expression for the disjoining pressure 
(\ref{eqn:generalpress}), leads to a very simple final result,
\be
\beta\Pi= 4n_0\sinh^2 \frac{\fm}2 ={n(x=0)}.
\ee
Although it might be tempting to attribute this simple final result to 
the contact-value theorem for charged plates, this is not the 
case (see discussion at the end of Section 3). 

\subsection{The overlapping regime}

For the overlapping regime, it is convenient to apply
the parametric differentiation just on the 
last term, $\bar{f}_{\mathrm{aux}}$, of the total free 
energy (\ref{eqn:freeover}), since the two first terms yield a 
constant contribution to the disjoining pressure,
\be
\frac{\beta\Pi}{n_0} \equiv \left. -\frac{\kappa}{n_0} 
\frac{\partial\bar{f}}{\partial\bar{h}}\right|_{\alpha,\bar{s}}
= 2\left(1-\e^{-4\alpha}\right)-4\left(1-\e^{-2\alpha}\right)  
\left. -\frac{\kappa}{n_0} 
\frac{\partial\bar{f}_{\mathrm{aux}}}{\partial\bar{h}}\right|_{\alpha,\bar{s}}.
\label{eqn:pressoverlap}
\ee
The derivatives of the last term (\ref{eqn:freeauxover}) of the total free energy,
$\bar{f}_{\mathrm{aux}}$, are given by  
\bea
\frac{\kappa}{n_0} \frac{\dd\bar{f}_{\mathrm{aux}}}{\dd\fm}&=&
4\e^{-2\alpha}{\cal E}_{2\alpha}(\fm,\fl) \sinh\left(\frac{\fm}2+2\alpha\right)\nonumber\\
&& -\frac{2\e^{-2\alpha}\sinh\left(\fm+4\alpha\right)}{\Delta_{2\alpha}(\fm,\fl)}
\left[\fl-2 \tanh^2\left(\frac{\fm}2+2\alpha\right) 
\coth\left(\frac{\fl}2+2\alpha \right)\right],\\
\frac{\kappa}{n_0} \frac{\dd\bar{f}_{\mathrm{aux}}}{\dd\fl}&=&
\frac{2\e^{-2\alpha}\sinh\left(\fl+4\alpha\right)}{\Delta_{2\alpha}(\fm,\fl)}
\left[\fl-2\tanh\left(\frac{\fl}2+2\alpha \right)\right] \nonumber\\
&&-\frac{ 2\e^{-\alpha} \sinh\left(\fl+2\alpha\right)}
{\Delta_\alpha(\fii,\fl)}\left[\fl -2\tanh\left(\frac{\fl}{2}+\alpha\right) \right],\quad\ \\
\frac{\kappa}{n_0} \frac{\dd\bar{f}_{\mathrm{aux}}}{\dd\fii}&=&-4\e^{-\alpha}
\left[{\cal E}_{\alpha}(\fii,\fl)-{\cal E}_{\alpha}(\fii,\fg)\right] 
\sinh\left(\frac{\fii}2+\alpha\right) \nonumber\\
&&+\frac{2\e^{-\alpha}\sinh\left(\fii+2\alpha\right)}{\Delta_\alpha(\fii,\fl)}
\left[\fl-2 \tanh^2\left(\frac{\fii}2+\alpha\right)
\coth\left(\frac{\fl}2+\alpha \right)\right] \nonumber\\
&&-\frac{2\e^{-\alpha}\sinh\left(\fii+2\alpha\right)}{\Delta_\alpha(\fii,\fg)}
\left[\fg-2 \tanh^2\left(\frac{\fii}2+\alpha\right)
\coth\left(\frac{\fg}2+\alpha \right)\right] , \\
\frac{\kappa}{n_0} \frac{\dd\bar{f}_{\mathrm{aux}}}{\dd\fg}&=&
\frac{2\e^{-\alpha} \sinh\left(\fg+2\alpha\right)}
{\Delta_{\alpha}(\fii,\fg)}\left[\fg -2\tanh\left(\frac{\fg}{2}+\alpha\right)
\right]+ 2\cosh\frac{\fg}2 \left(\fg-2 \tanh\frac{\fg}2\right)
 .\qquad
\eea

Now the defining equations for $\bar{h},\bar{s},u$ and $v$ are 
\bea
\bar{h}&=& 
\frac{{\cal F}_{\alpha}(\fii,\fg) -
{\cal F}_{\alpha}(\fii,\fl)}
{\e^{-\alpha} \cosh\left(\frac{\fii}{2} +\alpha\right)}, \\
\bar{s}&=& \frac{{\cal F}_{2\alpha}\left(\fm,\fl\right)}
{\e^{-2\alpha}\cosh\left(\frac{\fm}{2}+2\alpha\right)}+
\frac{{\cal F}_{\alpha}(\fii,\fg) -
{\cal F}_{\alpha}(\fii,\fl)}
{2\e^{-\alpha}\cosh\left(\frac{\fii}{2} +\alpha\right)},\\
u&=&\e^{-\alpha}\Delta_\alpha(\fii,\fl)-\e^{-2\alpha}\Delta_{2\alpha}(\fm,\fl)=0,\\
v&=&\e^{-\alpha}\Delta_\alpha(\fii,\fg)+2\sinh\frac{\fg}2=0,
\eea
and their associated derivatives for the evaluation of the Jacobian
${\partial(\bar{h},\bar{s},u,v)}/{\partial(\fm,\fl,\fii,\fg)}$,
\bea
\frac{\dd\bar{h}}{\dd\fm} &=& 0,\\
\frac{\dd\bar{h}}{\dd\fl} &=& -\frac1{\e^{-\alpha}\Delta_\alpha(\fii,\fl)}, \\
\frac{\dd\bar{h}}{\dd\fii} &=& 
\frac{\left[{\cal E}_{\alpha}(\fii,\fl)-{\cal E}_{\alpha}(\fii,\fg)\right]}
{2\e^{-\alpha}\sinh\left(\frac{\fii}2+\alpha\right)}
+\left[\frac{\coth\left(\frac{\fl}2+\alpha\right)}{\e^{-\alpha}\Delta_\alpha(\fii,\fl)}-
\frac{\coth\left(\frac{\fg}2+\alpha\right)}
{\e^{-\alpha}\Delta_\alpha(\fii,\fg)}\right]
\tanh\left(\frac{\fii}2+\alpha\right) ,\quad \\ 
\frac{\dd\bar{h}}{\dd\fg} &=& \frac1{\e^{-\alpha}\Delta_\alpha(\fii,\fg)}, \\
\frac{\dd\bar{s}}{\dd\fm}&=& 
-\frac{{\cal E}_{2\alpha}\left(\fm,\fl\right)}{2\e^{-2\alpha}\sinh\left(\frac{\fm}2+2\alpha\right)}
-\frac{\tanh\left(\frac{\fm}2+2\alpha\right)\coth\left(\frac{\fl}2+2\alpha\right)}
{\e^{-2\alpha}\Delta_{2\alpha}(\fm,\fl)}, \\
\frac{\dd\bar{s}}{\dd\fl}&=& \frac1{\e^{-2\alpha}\Delta_{2\alpha}(\fm,\fl)} -
\frac1{2\e^{-\alpha}\Delta_\alpha(\fii,\fl)} , \\
\frac{\dd\bar{s}}{\dd\fii}&=&
\frac{\left[{\cal E}_{\alpha}(\fii,\fl)-{\cal E}_{\alpha}(\fii,\fg)\right]}
{4\e^{-\alpha}\sinh\left(\frac{\fii}2+\alpha\right)}
+\left[ \frac{\coth\left(\frac{\fl}2+\alpha\right)}{2\e^{-\alpha}\Delta_\alpha(\fii,\fl)}
- \frac{\coth\left(\frac{\fg}2+\alpha\right)}
{2\e^{-\alpha}\Delta_\alpha(\fii,\fg)}\right] 
\tanh\left(\frac{\fii}2+\alpha\right) ,\qquad \\
\frac{\dd\bar{s}}{\dd\fg} &=& \frac1{2\e^{-\alpha}\Delta_\alpha(\fii,\fg)}, \\
\frac{\dd u}{\dd\fm} &=& \frac{\e^{-2\alpha}\sinh\left(\fm+4\alpha\right)}
{\Delta_{2\alpha}(\fm,\fl)}, \\
\frac{\dd u}{\dd\fl} &=& -\frac{\e^{-2\alpha}\sinh\left(\fl+4\alpha\right)}
{\Delta_{2\alpha}(\fm,\fl)}+
\frac{\e^{-\alpha}\sinh\left(\fl+2\alpha\right)} {\Delta_\alpha(\fii,\fl)}, \\
\frac{\dd u}{\dd\fii} &=& -\frac{\e^{-\alpha}\sinh\left(\fii+2\alpha\right)}
{\Delta_\alpha(\fii,\fl)}, \\
\frac{\dd u}{\dd\fg} &=& \frac{\dd v}{\dd\fm} = \frac{\dd v}{\dd\fl} = 0,\\
\frac{\dd v}{\dd\fii} &=& 
-\frac{\e^{-\alpha}\sinh\left(\fii+2\alpha\right)} {\Delta_\alpha(\fii,\fg)}, \\
\frac{\dd v}{\dd\fg} &=&
\frac{\e^{-\alpha}\sinh\left(\fg+2\alpha\right)}{\Delta_\alpha(\fii,\fg)}
+ \cosh\frac{\fg}2 .
\eea

Putting all together into the disjoining pressure expression
(\ref{eqn:pressoverlap}), we obtain for the overlapping regime, 
\bea
\beta\Pi
&=&n_0 \left( 
\e^{-\fm-8\alpha} + \e^{\fm} -2 + 2\e^{-\fl-4\alpha}-2\e^{-\fl-8\alpha} \right)
\nonumber\\
&=&2n_0\left[\e^{-4\alpha}\cosh\left(\fm+4\alpha \right)-1+2\e^{-\fl-6\alpha}
\sinh 2\alpha\right] \nonumber\\
&=& n(x=0)+ 2\Delta n_+(x=s-\frac{h}2)  ,
\eea
where the discontinuity of the density of cations $\Delta n_+$ upon
crossing the surface located at $x=s-\frac{h}2$ is defined 
by~(\ref{eqn:nplusdisc}).
We remark that, due to this additional term, in this case the disjoining pressure 
\textit{has not the form} of the expression given by the contact-value theorem for
charged plates. 

\section*{Acknowledgments}

We acknowledge fruitful discussions with M. C. Barbosa, 
A. W. C. Lau, Y. Levin and
E. M. Mateescu.
This work was supported by the Brazilian agency 
CNPq --- Conselho Nacional de
Desenvolvimento Cient{\'\i}fico e Tecnol{\'o}gico.
This research was also partially supported by the MRL Program 
of the National Science Foundation under Award Nos.~DMR-96-32716,
DMR-96-24091 and DMR-97-08646.

\end{document}